\documentstyle[11pt]{article}
\title{Exact solutions for a mean-field Abelian sandpile\thanks{
                Supported in part by NSF Grant DMR89-18903}}
\author{Steven A. Janowsky\thanks{Supported in part by NSF Mathematical
                Sciences Postdoctoral Research Fellowship DMS
                90-07206}\hspace{.25em}\thanks{Address after August 1993:
		Department of Mathematics, University of Texas,
		Austin, TX 78712}\\
	Department of Mathematics\\
        Rutgers University\\New Brunswick, NJ 08903
\and
        Claude A. Laberge\\
        Department of Physics\\
        Rutgers University\\Piscataway, NJ 08855-0849}
\date{June 1993}
\begin{document}
\maketitle
\renewcommand{\baselinestretch}{1.57}\large\normalsize
\begin{abstract}
We introduce a model for a sandpile, with $N$ sites, critical height $N$
and each site connected to every other site.  It is thus a mean-field
model in the spin-glass sense.  We find an exact solution for the
steady state probability distribution of avalanche sizes, and discuss
its asymptotics for large $N$.
\end{abstract}

The publication of a series of articles by P. Bak and collaborators
\cite{Bak1,Bak2} generated a growing interest in the study of certain
class of cellular automata models now commonly known as ``sandpiles''
because of a crude analogy between the dynamical rules and the way sand
topples when building a sandpile. The basic motive was not to model the
way real sandpiles behave, although it did broaden the audience for
granular flow (see e.g.~\cite{Nag,Mehta,Puhl}), but to study a
phenomenon they coined ``self-organized criticality'' in which a
system reaches a critical
state, i.e.\ a state with no intrinsic length or time scales, without
the tuning of an external parameter. This generated a big influx of
theoretical, computational and even experimental studies of
self-organized criticality and of sandpiles in general.  Unfortunately,
even though the number of models (i.e.\ sets of dynamical rules)
exhibiting self-organized criticality or falling into the more general
category of sandpiles has increased dramatically, very few exact results
are known at the present (see e.g.\ \cite{DM1,DM2,DM3,DR,Lee}).
In this letter we present a model in which
many quantities can be calculated exactly in a rigorous way.


A sandpile model is basically a set of dynamical rules describing the
way that grains of sand are added to a system, the conditions under
which those grains can be redistributed inside the system, and the way
they are removed from the system. Here we consider a system of $N$ sites
and define $h(i)$ as the (integer) height of the column of sand at site
$i$, $i\in \{1\ldots N\}$. We drop a grain of sand on a site $i$ chosen
at random, thereby increasing its height by one: $h(i) \rightarrow
h(i)+1$. If this new height exceeds the maximum stable value $h_c$ then
that column topples and gives 1 grain of sand to each of the $N-1$ other
sites while one grains drops out of the system.  (We take $h_c \ge N$ so
that $h(i)\ge 0$; in fact we are primarily interested in $h_c=N$.) We
then examine the system to see if any site has a column exceeding $h_c$
in which case we topple that column also.  We keep toppling until all
the sites are stable (this characterizes an avalanche).  We then repeat
the procedure of adding a grain at a randomly chosen site.

This model falls into the category of abelian sandpiles since we always
obtain the same stable configuration from an unstable one irrespective
of the order in which we executed the topplings. Dhar~\cite{Dhar} used
this fact to obtain several properties of those models which we use
extensively in our analysis. Furthermore, since all sites are connected
to all other sites, we can say that we have a mean field theory of
abelian sandpiles. It is this combination that permits the model to be
solved. It should be pointed out that our model is quite different from
previous attempts to study mean-field sandpiles \cite{Tang,Gaveau}.


It is shown in~\cite{Dhar} that not all configurations are allowed in the
asymptotic regime, but that the allowed (recurrent) configurations all
have equal weight. In general, the number of recurrent configuration is
given by the determinant of the toppling matrix $\Delta$ which is an $N
\times N$ matrix where $\Delta_{ii}=h_c$ and $\Delta_{ij} = -1$ for each
site $j$ connected to $i$, so that row $i$ of $\Delta$ represents the
amount of sand lost by every site when site $i$ topples. Thus in the
model considered here this matrix takes the form
\begin{equation}
\Delta_{ij}= -1 + (h_c + 1)\delta_{ij},
\end{equation}
where $\delta_{ij}$ is the Kronecker delta.

Because of the highly symmetric nature of $\Delta$ in our model
it is straightforward to calculate the determinant and determine that
the number of recurrent configurations is
\begin{equation}
Z(N, h_c) = \det\Delta = (h_c + 1 - N)(h_c + 1)^{N-1}.
\end{equation}
For the case $h_c=N$ we have $Z(N,N)=(N+1)^{N-1}$. Incidentally this is
the number of spanning trees on a fully connected graph with $N+1$
sites. Such a result was expected from the general observation that
$\det\Delta$ is precisely the cofactor of the matrix tree (see e.g.\
\cite{Graph}) for graphs defined on a superset of the sandpile lattice
(there is an additional site, the ground, connected to every site). In
fact there is a one-to-one correspondence between the recurrent
configurations and the spanning trees, in the sense that a configuration
is recurrent if and only if it passes the ``burning algorithm'' test
described in \cite{DM1,DM3}. Actually this holds only for symmetric abelian
sandpiles, i.e.\ models where $\Delta$ is symmetric; for the asymmetric
case see \cite{Speer}.

Clearly the recurrent configurations fall into equivalent classes, where
two configurations are equivalent if one is a permutation of the other.
Because of the symmetry of the model we need only examine one member of
each equivalence class; we find it convenient to restrict ourselves to
the case $h(i)\le h(i+1)$, $i=1\ldots N-1$, where the amount of sand
increases as the site label increases.

Our analysis will focus on a particular subset of the recurrent
configurations, namely the minimal configurations.  A given
configuration is minimal if there is no recurrent configuration that can
be obtained by removing sand from the given configuration.  Some brief
computations show (one can use the so-called ``burning algorithm''
of~\cite{DM1}) that the configuration $h(i) = i$ is a minimal
configuration.  All minimal configurations fall into this equivalence
class and thus there are exactly $N!$ minimal configurations, the
permutations of $h(i) = i$.  If we now consider equivalence classes for
general configurations, we see that the restriction that a recurrent
configuration be greater than or equal to the minimal configuration
yields
\begin{equation}\label{stability}
h(i) \ge i, \qquad
h(i) \ge h(i-1)
\end{equation}
(set $h(0) = 0$).

Now consider how avalanches of various sizes come about in our model; we
limit the discussion to the case $h_c=N$.  Size here means either the
number of grains that fall out of the system or the number of sites that
topple; in the model considered here there is no ambiguity because they
are equal.  An avalanche begins when a grain of sand is dropped on a
site of height $N$.  It will continue if the second-highest site is also
at height $N$.  It will be at least of size three if the third-highest
site is at height $N-1$, and it will be at least of size four if the
 fourth-highest site is at height $N-2$, etc.  Therefore the size will be
(using the equivalence class notation)
\begin{equation}\label{N-aval}
N_{\rm aval}= N- \min \{j: h(i) > i {\rm\ for\ all\ } j\le i<N\}.
\end{equation}
This is the only non-zero size avalanche this configuration can produce,
and it is determined simply by ``how long'' the configuration stays
away from a minimal one.

If we have such a configuration, what is the probability of starting an
avalanche? It is simply the fraction of sites at the maximum value,
$|\{i: h(i) = N\}|/N$.  Therefore, to determine the avalanche size
probability distribution, we merely have to count the configurations
with fixed $N_{\rm aval}$ and with a fixed number of sites at height $N$.
Luckily, the number of such configurations can be simply represented as
partition functions for smaller systems with different critical
heights---we obtain (for $1<k\le N$):
\begin{eqnarray}
\lefteqn{P({\rm avalanche\ of\ size\ }k) =}
                \nonumber\\
&&Z(N,N)^{-1}Z(N-k, N-k) \sum_{j=2}^k {N \choose j, k-j, N-k}
        \frac{j}{N} Z(k-j,k-2).         \label{avalanchesum}
\end{eqnarray}

A brief explanation of the source of the terms in (\ref{avalanchesum})
is in order.  First of all, since site $N-k$ must be at height $N-k$ to
get an avalanche of the right size, the remaining sites below $N-k$
yield a factor $Z(N-k, N-k)$.  Next $j$ is the number of sites at height
$N$.  The sites between $N-k+1$ and $N-j$ must be greater than minimal yet
less than $N$; this is equivalent to $Z(k-j,k-2)$. The combinatoric
 factor gives us the number of ways we can choose configurations with
 fixed equivalence classes for the subsystems.  Finally we have the
probability of toppling $j/N$ and the normalization $Z(N,N)$.

Simpler expressions hold for $k=0,1$:
\begin{equation}
P({\rm avalanche\ of\ size\ }1) = Z(N-1, N-1)/Z(N,N);
\end{equation}
\begin{equation}\label{noavalanche1}
P({\rm no\ avalanche)} = \sum_{j=1}^N {N \choose j} \frac{N-j}{N}
Z(N-j,N-1)/Z(N,N).
\end{equation}
The above sums (\ref{avalanchesum}--\ref{noavalanche1}) can be performed
exactly.  The result is
\begin{equation}\label{exactav}
P({\rm avalanche\ of\ size\ }k) =
\frac{(N-k+1)^{N-k-1} k^{k-2}(N-1)!}{(N+1)^{N-1} (k-1)! (N-k)!}
\end{equation}
 for $1\le k\le N$, and
\begin{equation}\label{noavalancheresult}
P({\rm no\ avalanche}) = \frac{N-1}{N+1}.
\end{equation}
The probability distribution (\ref{exactav}) has the unusual feature
that (for $k\ne 0$) it is symmetric about $k=(N+1)/2$.

We are interested in the large $N$ behavior of this result.  Taking
$k>0$ fixed and $N\rightarrow\infty$, (\ref{exactav}) yields
\begin{equation}
P({\rm avalanche\ of\ size\ }k) \sim \frac1N\frac{k^{k-2}e^{-k}}{(k-1)!},
\end{equation}
and if k is also large ($1\ll k\ll N$) then this reduces to
\begin{equation}
P({\rm avalanche\ of\ size\ }k) \sim \frac{k^{-3/2}}{\sqrt{2\pi}N}.
\end{equation}
Thus we reproduce the exponent $-3/2$ previously derived on
trees~\cite{DM2} and via numerical and nonrigorous
arguments~\cite{Obukov,KNWZ}.


Another quantity of interest is the single-site probability distribution
for the heights, {\em i.e.}\ the probability that a given site contains
$H$ grains of sand in the stationary state.  First look at
configurations where exactly $K$ sites have height $H$.  (In the
equivalence class notation these sites must be consecutive.)  Remove
these sites and consider the subsystem consisting of the remaining $N-K$
sites.  The recurrent configurations of this subsystem are in one-to-one
correspondence with those of a system of size $N-K$ where the recurrent
configurations are determined by, in addition to the usual occupancy
restrictions (see (\ref{stability})), the conditions $j \le h(j) \le
N-1$ if $j \le H-K$ and $j+K-1 \le h(j)\le N-1$ if $j > H-K$.

The direct evaluation of the number of allowed configurations satisfying
the above criteria is possible but somewhat complicated; appropriately
summing over $K$ would yield the desired probabilities.  However, it
turns out to be simpler instead to compare the number of allowed
configurations with $K$ sites having $H$ grains with the number of
allowed configurations with $K$ sites having $H-1$ grains, {\em e.g.}
by subtracting the latter from the prior.  Which configurations are left?

In both cases $j \le h(j) \le N-1$ for $j < H-K$ and $j+K-1 \le h(j)\le
N-1$ if $j > H-K$.  The only difference occurs at site $H-K$ which can
be occupied by between $H-1$ and $N-1$ grains in the first case, and
between $H-K$ and $N-1$ grains in the second. If $K=1$ these conditions
are identical and no allowed configurations are left. For $K \geq 2$,
however, site $H-K$ is now restricted so that $h(H-K) \le H-2$.  Since
we need $h(j+1) \geq h(j)$ (for the configuration to be a recurrent
one), this implies that the first $H-K$ sites can have a maximum of
$H-2$ grains each and thus contribute a factor $Z(H-K,H-2)$ to the
partition function, excluding (global) symmetry factors.  On the other
hand, (\ref{stability}) imposes no additional conditions on the last
$N-H$ sites, and it is easily seen that they contribute a factor
$Z(N-H,N-H)$.  Defining $P(H)\;$= the probability that a given site will
have $H$ grains, putting in the appropriate symmetry factors and summing
over the index $K$ we get
\begin{equation}
P(H) - P(H-1) = Z(N,N)^{-1}Z(N-H, N-H) \sum_{K=2}^H {N \choose K, H-K, N-H}
        \frac{K}{N} Z(H-K,H-2),
\end{equation}
which is exactly (\ref{avalanchesum})! Since $P(0)=0$ we can collapse
the telescoping sum and obtain
\begin{equation}\label{height-dist}
P(\mbox{site has $H$ grains}) = \sum_{k=1}^H P(\mbox{avalanche of size k}).
\end{equation}
Because of the symmetry of the avalanche distribution, we have here the
symmetry
\begin{equation}
P(\mbox{site has $H$ grains}) + P(\mbox{site has $N-H$ grains}) =
\frac{2}{N+1}
\end{equation}
for $0\le H \le N$.

The asymptotics of (\ref{height-dist}) are relatively easy to compute using our
previous results.  For $N\gg1$ we have
\begin{equation}
P(\mbox{site has $H$ grains}) \sim \frac1N\sum_{k=1}^H
\frac{k^{k-2}e^{-k}}{(k-1)!}
\end{equation}
which can be used directly to compute the behavior for small $H$.
For $N\gg H\gg 1$ we have
\begin{equation}
P(\mbox{site has $H$ grains}) \sim \frac1{N+1}\left(1-\sqrt{\frac2{\pi
H}}\right).
\end{equation}


We can also examine the total amount of sand in the system.  The
distribution of masses (total number of grains of sand) of recurrent
configurations is most easily computed for mass $M$ near the maximum
value $N^2$.  For example, the number of recurrent configurations with
mass $N^2-2$ is simply $N(N+1)/2$.  This is easily generalized to
arbitrary mass (using inclusion-exclusion arguments) and we find that
the number of recurrent configurations with total mass $N^2-K$,
$Z_{N^2-K}(N,N)$, is
\begin{eqnarray}
\lefteqn{Z_{N^2-K}(N,N) = {N-1+K \choose N-1} +}\nonumber\\
&&\sum_{l=1}^{l(N,K)} \!(-1)^l {N \choose l}
        \sum_{j=lN+l(l-1)/2}^K {N-l-1+K-j \choose N-l-1}
        {j - lN - l(l+1)/2 -1 \choose l-1},
        \label{massdist}\end{eqnarray}
where
\begin{equation}
l(N,K) = \left\lceil N - (N(N-5) + 2K + 17/4)^{1/2} - 3/2
\right\rceil
\end{equation}
simply indicates ``how far'' one must go in the inclusion-exclusion;
$\lceil A \rceil$ is the least integer $\ge A$.  Then the probability of
a configuration having mass $M$ is $Z_M(N,N) / (N+1)^{N-1}$.
Unfortunately we are only able to find the asymptotic behavior of
(\ref{massdist}) for the tails of the distribution, i.e.\ $M$ near
$N(N+1)/2$ or $M$ near $N^2$, which is not where most of the (recurrent)
configurations reside.  Expressions similar to that which were used
to find the avalanche distribution, e.g.\ (\ref{avalanchesum}), are
available:
\begin{equation}
Z_M(N,h_c) = \sum_{j=1}^N Z_{M-jh_c} (N-j, h_c-1) {N \choose j},
\end{equation}
and one could consider a generating function approach to eliminate the
restrictions on $j$ in the sum.  Unfortunately the resulting expression
will only converge in the regions equivalent to very large or very small
mass, and thus provide no new information.


The reformulation of the $N$-dimensional model in terms of a one
dimensional sandpile-like problem (with the required symmetry factor
attached to each configuration) greatly facilitated the computations of
``static'' quantities like the avalanche size, single-site height, and
total mass distributions. In order to study the ``dynamical'' quantities
associated to the evolution of an avalanche, another reformulation in
term of a one dimensional {\em particle} model proves to be useful; we
will use it to calculate the distribution of avalanche durations.  This
reformulation also provides a much more efficient method for numerical
study than a straightforward implementation of the sandpile process.

The duration of an avalanche is the number of sweeps that are required
in order to reach a stable configuration once a grain of sand is dropped
on the system.  A sweep consists of two steps: we first go through all
the sites and mark those with a number of grains exceeding the critical
value and then topple all those sites simultaneously. The process is
repeated until no site can topple.  Note that this definition of
duration is not the only one; others are possible because of the abelian
nature of the model.

The particle model consists of $N$ particles moving on a one dimensional
lattice with sites labeled from left to right by $k \in [1,N+1]$ (where
N is the number of site in the original model).  The correspondence
between sandpile configurations and particle configurations is as
follows: given a sandpile configuration $C_0$ place a particle at
position k for every site in $C_0$ containing k grains.  If $n_k$ is the
resulting number of particles at site $k$, the stability condition
(\ref{stability}) translates to $n_k \leq k$ and $n_N >1$; since the
critical height is $N$ we also have $n_{N+1}=0$.

The boundary conditions are ``almost'' periodic as will be explained
below; we also introduce a marker (barrier) between sites $N$ and $N+1$
that will play an important role in the dynamics.

The dynamics of the particle model is as follow: pick a particle at
random and move it one site to the right (this is equivalent of dropping
a grain of a sand on $C_0$). If the barrier was not crossed then we have
a new stable configuration and can repeat the step.  Thus, neglecting
the boundary, our model is a zero-range process; however, the boundary
plays a very important role.

If the chosen particle jumps through the barrier it means that one site
in $C_0$ is unstable and will topple. In the particle model such a
toppling will mean that all particles simultaneously jump one site to
the right except for the one at site $N+1$ which should move back to
site 1.  An equivalent formulation which we find more convenient is to
move the barrier to the left by one unit and relabel the sites
accordingly (i.e.\ the barrier defines the position of site N and N+1 on
the periodic lattice).

As we moved the barrier, a number of particles (say $p_2$) might have gone
through it.  In the sandpile model this means that the initial toppling
caused $p_2$ more sites to become unstable. To duplicate their toppling
the barrier should now be moved $p_2$ more units to the left. The process
continues until no particle crosses the barrier as it jumps. The total
number of barrier jumps required is the duration of the avalanche.
The probability that an avalanche has duration $T$ is equivalent to the
fraction of all particle configurations that will make the barrier jump
$T$ times if a particle at site $N$ is picked.  Of course $T=0$
corresponds to the no avalanche case so that the probability that the
avalanche duration is zero is $(N-1)/(N+1)$ (see (\ref{noavalancheresult})).

If there are $T$ barrier jumps, let $p_1, p_2, \ldots p_T$ be the number of
particles crossing the boundary at each jump and let $P = p_1 + p_2 +
\cdots + p_T$; note that $p_i>0$ for $i<T$ and
$p_T=0$; we also have $p_1 > 1$ for $T>1$.  Provided $T>0$ we have
\begin{equation}
P(\mbox{duration $T$})=Z(N,N)^{-1}\sum_{p_1,\ldots,p_T\rm~restricted}
\frac{p_1}{N}W_N(1; p_1, p_2, \ldots, p_T)
\end{equation}
where $W_N(1; p_1, p_2, \ldots, p_T)$ is the number of configurations
that produce the specified barrier jumps.  We can compute this in an
iterative fashion, in terms of $W_{M}(s; p_i, p_{i+1}, \ldots, p_T)$, the
number of (restricted) configurations on the subsystem consisting of
sites $1\ldots M$; $s$ is size of the current jump, and $p_i, p_{i+1},
\ldots, p_T$ are the remaining jump sizes.  We have
\begin{eqnarray}
W_N(1; p_1, p_2, \ldots, p_T) &=& {N \choose p_1} w(p_1,1)
	W_{N-1}(p_1-1; p_2, p_3, \ldots, p_T),\\
W_{N-1}(p_1-1; p_2, p_3, \ldots, p_T) &=& {N-p_1\choose p_2} w(p_2,p_1-1)
	W_{N-p_1}(p_2; p_3, p_4, \ldots, p_T),\\
&&\mbox{and}\nonumber\\
W_{M}(p_{i-1}; p_i, p_{i+1}, \ldots, p_T) &=& {M \choose p_i} w(p_i,p_{i-1})
	W_{M-p_{i-1}}(p_i; p_{i+1}, p_{i+2}, \ldots, p_T)
\end{eqnarray}
for $2<i<T-1$.  The weight $w(p,s)$ is easily computed since the
criteria for allowed configurations are automatically satisfied within
each block.  Since each of the $p$ particles can independently be at any
of the $s$ sites,
\begin{equation}
w(p,s) = s^p.
\end{equation}
Finally, the last step of the iteration is simply
\begin{equation}
W_{N-P+p_{T-1}}(p_{T-1}; p_T) = {N-P+p_{T-1} \choose p_T} w(p_T,p_{T-1})
	Z(N-P,N-P) = Z(N-P,N-P).
\end{equation}
As an example, for $T=4$ this would lead to
\begin{equation}
P(\mbox{duration 4})= \sum_{p_1=2}^{N-2} \sum_{p_2=1}^{N-p_1-1}
\sum_{p_3=1}^{N-p_1-p_2} \frac{Z(N-P,N-P)}{Z(N,N)}\frac{p_1}{N}
{N\choose p_1\, p_2\, p_3\, p_4} (p_1-1)^{p_2} p_2^{p_3}.
\end{equation}

The general case can actually be written in a reasonably compact form.
We can rearrange the terms and obtain the following (for $T>0$):
\begin{equation}
P(\mbox{duration $T$})=\frac{1}{(N+1)^{N-1}}\!\! \sum_{S=T}^N \frac{1}{N}
{N \choose S}(N-S+1)^{N-S-1}
\!\!\!\!\!\!\!\sum_{s_1 +s_2 +\cdots+ s_T=S}  \frac{S!}{s_1!..s_T!}
s_1^{s_2} \ldots s_{T-1}^{s_T}
\end{equation}
where we have the restriction $s_1=1$, $s_i > 0$.
We identify $s_i$ with $p_{i-1}$ except that because of the need to
initiate the avalanche $s_1=1$ and $s_2 = p_1-1$. Of course $s_i$ is
nothing more than the size of the $i^{\rm th}$ jump.


\subsubsection*{Acknowledgments:}
We would like to thank J. L. Lebowitz, S. Majumdar and  E. R. Speer  for
their help with this work.

\renewcommand{\baselinestretch}{1.1}\large

\end{document}